\documentclass[x11names]{article}

\usepackage[nonatbib, preprint]{neurips_2024}

\usepackage[utf8]{inputenc} 
\usepackage[T1]{fontenc}    

\usepackage{xcolor}
\usepackage{color}
\usepackage{colortbl}
\definecolor{darkblue}{rgb}{0, 0, 0.5}
\usepackage[colorlinks=true, citecolor=darkblue, linkcolor=darkblue, urlcolor=darkblue]{hyperref}    
\usepackage{booktabs}       
\usepackage{amsfonts}       
\usepackage{nicefrac}       
\usepackage{microtype}      

\usepackage[numbers]{natbib}
\usepackage{times}
\usepackage{soul}
\usepackage{url}
\usepackage{graphicx}
\usepackage{amsmath}
\usepackage{amsthm}
\usepackage{amssymb}
\usepackage{amsfonts}
\usepackage[ruled,vlined]{algorithm2e}
\usepackage[tableposition=top]{caption}
\usepackage{multirow}
\usepackage{makecell}
\usepackage{booktabs}
\usepackage{enumitem}
\usepackage{footmisc}
\usepackage{subfigure}
\usepackage{appendix}
\usepackage{tabularx}
\usepackage{multibib}
\usepackage{wrapfig}
\newcites{apndx}{References in Appendix}
\usepackage{inconsolata}

\newif\ifpeerreview
\peerreviewfalse

\usepackage{xspace}
\def\ours{ASI++\xspace}
\usepackage{marvosym}

\title{ASI++: Towards Distributionally Balanced End-to-End Generative Retrieval}

\author{Yuxuan Liu\textsuperscript{$1\dagger$}\thanks{Contributed during internship at Microsoft.  \\ \indent\  \textsuperscript{$\dagger$}The authors contribute equally.}, Tianchi Yang\textsuperscript{$2\dagger$}, Zihan Zhang\textsuperscript{$2\dagger$}\textsuperscript{$\textrm{\Letter}$}, Minghui Song\textsuperscript{$2$}, Haizhen Huang\textsuperscript{$2$}, \\
\textbf{Weiwei Deng\textsuperscript{$2$}, Feng Sun\textsuperscript{$2$}, Qi Zhang\textsuperscript{$2$}}
\\\\
\textsuperscript{$1$}Peking University \quad \textsuperscript{$2$}Microsoft AI \\
{ yx.liu@stu.pku.edu.cn} \\ { \{tianchiyang,zihzha,minghuisong,hhuang,dedeng,sunfeng,zhang.qi\}@microsoft.com} 
}

\begin{document}
\maketitle

\begin{abstract}
Generative retrieval, a promising new paradigm in information retrieval, employs a seq2seq model to encode document features into parameters and decode relevant document identifiers (IDs) based on search queries. Existing generative retrieval solutions typically rely on a preprocessing stage to pre-define document IDs, which can suffer from a semantic gap between these IDs and the retrieval task. However, end-to-end training for both ID assignments and retrieval tasks is challenging due to the long-tailed distribution characteristics of real-world data, resulting in inefficient and unbalanced ID space utilization. To address these issues, we propose \ours, a novel fully end-to-end generative retrieval method that aims to simultaneously learn balanced ID assignments and improve retrieval performance. \ours builds on the fully end-to-end training framework of vanilla ASI and introduces several key innovations. First, a distributionally balanced criterion addresses the imbalance in ID assignments, promoting more efficient utilization of the ID space. Next, a representation bottleneck criterion enhances dense representations to alleviate bottlenecks in learning ID assignments. Finally, an information consistency criterion integrates these processes into a joint optimization framework grounded in information theory. We further explore various module structures for learning ID assignments, including neural quantization, differentiable product quantization, and residual quantization. Extensive experiments on both public and industrial datasets demonstrate the effectiveness of \ours in improving retrieval performance and achieving balanced ID assignments.
\end{abstract}

\section{Introduction}
Information retrieval plays an essential role in search engines, question-answering systems, and so on \cite{salton1991developments}. Recently, Generative Retrieval, a new end-to-end retrieval paradigm, has garnered significant attention \cite{sun2024learninggenret, tay2022transformerdsi, bevilacqua2022autoregressiveseal, yang2023auto}. This approach utilizes a seq2seq model to encode document features via model parameters and decode relevant document identifiers (IDs) in an autoregressive manner.  

Previous solutions in generative retrieval rely on a preprocessing stage to pre-define the document IDs as an atomic integer~\cite{zhou2022ultron, tay2022transformerdsi}, n-grams~\cite{bevilacqua2022autoregressiveseal, tang2023semanticsedsi, li2023multiviewminder, chen2022corpusbrain, chen2023unified}, natural language sentences~\cite{de2020autoregressiveentity, chen2022corpusbrain, chen2023unified}, hierarchical clustering~\cite{tay2022transformerdsi,wang2022neuralnci, zhuang2022bridgingdsiqg}, etc. These methods either suffer from the extensive time costs of preprocessing or face limitations due to the semantic gap between the preprocessing of ID assignment and the training of retrieval objectives.

To overcome these challenges, \citet{yang2023auto} developed a fully end-to-end generative retrieval framework ASI. ASI introduces a semantic indexing module that integrates ID assignments into the generative training framework. Specifically, ASI learns to map documents into discrete numeric IDs via the semantic indexing module and maps search queries to the IDs corresponding to relevant documents via an encoder-decoder model, combining these processes into a single generative retrieval training framework. 

However, end-to-end training for both ID assignments and retrieval tasks is complex due to the long-tailed distribution characteristics of real-world data, which lead to inefficient and unbalanced utilization of the ID space. This imbalance results in some IDs pointing to too many documents while others point to none, creating intervals that are either too dense or too sparse. Dense intervals can obscure fine-grained differences between documents and allow low-quality noise documents to interfere with the retrieval results, negatively impacting the accuracy and relevance of the documents that the IDs should correspond to.

To overcome the limitations above, we propose \ours, a novel fully end-to-end generative retrieval method designed to simultaneously learn a distributionally balanced ID assignment and improve document retrieval. \ours builds upon ASI, retaining its advantages such as fully end-to-end training, efficient retrieval based on one-to-many ID assignments, and generalization capabilities to new documents. To further address the issues of inefficient and unbalanced ID space utilization, \ours incorporates several innovative criterions. 

First, we propose a distributionally balanced criterion for ID assignments, promoting a balanced utilization of ID space by spreading learned IDs more evenly. Considering the dense representations from the encoder form a representation bottleneck, we introduce a representation bottleneck criterion to make it easier for the subsequent semantic indexing module to learn efficient and balanced ID assignments. Additionally, we integrate these optimization processes from an information theory perspective, jointly optimizing both parts through an information consistency criterion.

Based on the above improvements, we further explore different structures of the semantic indexing module, including differentiable Product Quantization (PQ)-based \cite{zhan2022repconc} and Residual Quantization (RQ)-based \cite{lee2022autoregressiverq, rajput2024recommenderrq} modules, and the MLP-based neural quantization module in ASI \cite{yang2023auto}.

Finally, we conduct extensive experiments on both large-scale public dataset and industrial dataset, which verifies that our \ours can effectively utilize the indexing space and improve ID assignments during end-to-end training, thereby outperforming all the competitive baseline and achieving SOTA performance. 
Our contributions are summarized as follows: 

\begin{itemize}
    \item 
    We propose \ours, a novel fully end-to-end generative retrieval method that simultaneously learns semantically balanced ID assignments and improves document retrieval. We also explore three different structures of the semantic indexing module in \ours.

    \item 
    To address the inefficient and unbalanced utilization of indexing space, \ours incorporates novel criteria that foster balanced utilization of ID space and enhance dense representations. These designs are integrated into a cohesive framework based on information theory.
    
    \item 
    Extensive experiments on public and large-scale industrial datasets demonstrate the effectiveness of \ours in terms of both retrieval performance and balancing ID assignments.
\end{itemize}

\section{Methods}
We present \ours, an end-to-end generative retrieval framework based on a transformer-based encoder-decoder equipped with a semantic indexing module.

To achieve a fully end-to-end retrieval, following \citet{yang2023auto}, \ours maps documents into discrete numeric IDs via the semantic indexing module, maps search queries into the IDs corresponding to the relevant documents via the encoder-decoder, and integrates these processes into a single generative retrieval training framework.

To enhance ID assignments while ensuring balanced indexing space utilization, we propose several novel semantic training criteria at both the discrete ID level and dense representation level. For discrete IDs, in addition to the discrete contrastive loss~\cite{yang2023auto}, we introduce a distribution balancing criterion, incorporating a novel distribution density loss to balance ID assignments using both Euclidean and exact match distances. At the dense representation level, we propose analogous contrastive and density losses, forming the representation bottleneck criterion. Finally, we design an information consistency criterion, implementing a semantic compression mechanism that jointly optimizes dense representations and discrete IDs from an information-theoretic perspective.

\begin{figure}[!t]
  \centerline{
  \includegraphics[width=0.98\linewidth]{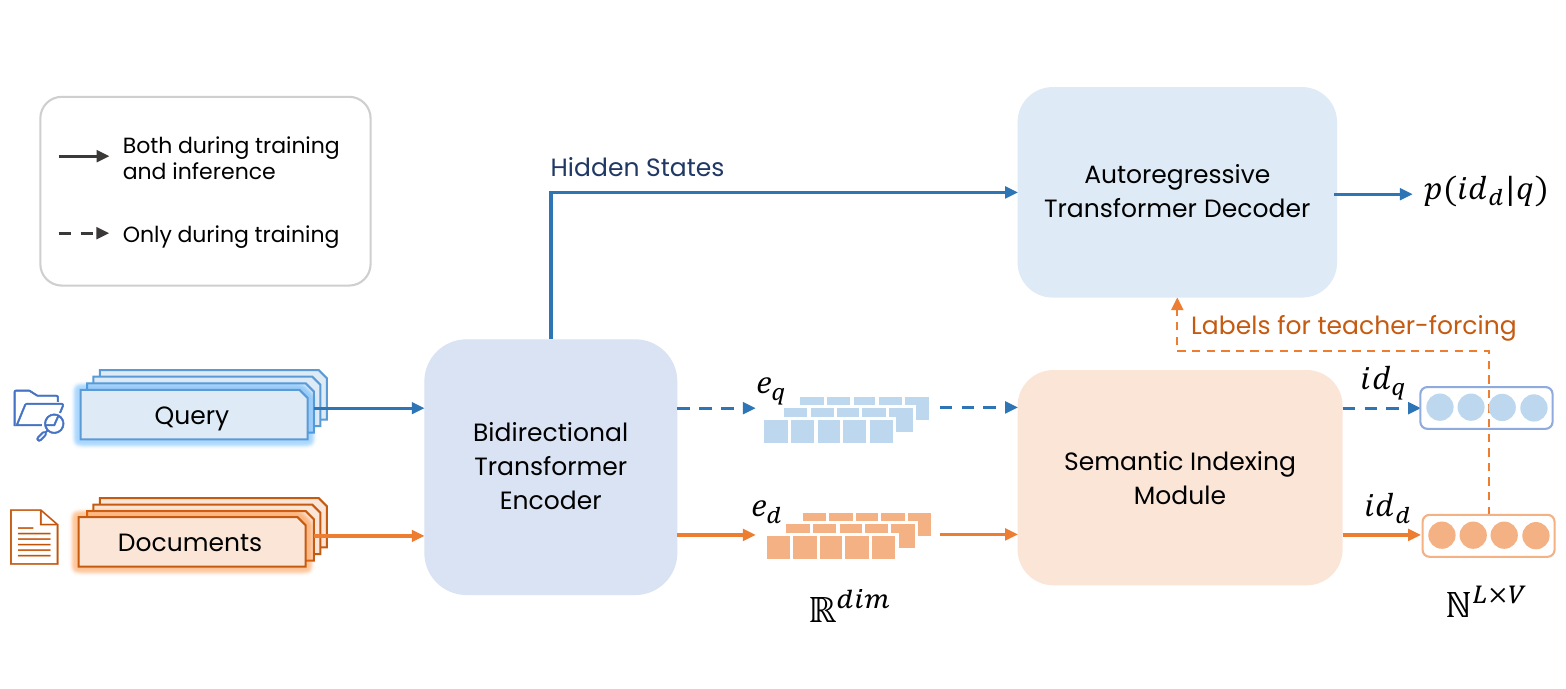}}
  \caption{Schematic diagram of training and inferring in our proposed \ours. Given a search query $q$ or document $d$, \ours first encodes them into dense latent representations via encoder: $e_q$, $e_d$. Then \ours obtains discrete numeric index $id_q$, $id_d$ via semantic indexing module or decoder, respectively. }
  \label{fig:model}
\end{figure}

\subsection{Model Architecture}
In this section, we introduce the architectural design of \ours. Following \citet{yang2023auto}, \ours consists of two main components, as illustrated in Figure \ref{fig:model}: an autoregressive Transformers \cite{vaswani2017attention}-based encoder-decoder and a semantic indexing module. 
Specifically, given a query-document pair $q$-$d$, the encoder captures their semantic information and maps them into a dense latent representation of dimension $ {dim} $: $ e_q, e_d \in \mathbb{R}^{dim}$. Then the decoder autoregressively predicts the discrete numeric ID $id_d \in \mathbb{N}^{L \times V}$ of the document $ d $, where the ID is automatically learned by the semantic indexing module based on its dense representation $ e_d $.

Formally, given a query-document pair $(q,d)$, the encoder first embeds them into dense representations: $e_q = \texttt{Encoder}(q); e_d = \texttt{Encoder}(d)$, and then the indexing module assigns them with discrete IDs via the distribution of discrete numeric index, as $p(id_q|e_q) = \texttt{Indexing}(e_q), id_q = \arg \max p(id_q|e_q)$ (take $q$ as an example). Following \cite{yang2023auto}, we employ a discrete contrastive (margin) loss to encode the retrieval supervision signals into the ID semantics as follows: 
\begin{equation}
    \mathcal{L}_c = \sum\nolimits_{(q,d) \in B, (q,d^-) \not\in B} \max\left[0, \mathfrak{D}(id_q, id_{d}), \mathfrak{D}(id_q, id_{d^-}) + \alpha\right] , 
\end{equation}
where $ \alpha $ denotes distance margin, $d^-$ denote in-batch negatives for query $q$, $\mathfrak{D}(\cdot, \cdot)$ denotes distance $\mathfrak{D}(id_q, id_{d}) = \sum\nolimits_i^L||p(id_{q_i}|e_q) - p(id_{d_i}|e_d)||^2$. 
Meanwhile, the encoder-decoder employs a teacher-forcing generation loss to learn the mapping from query $q$ to the IDs $id_d$ corresponding to document $d$ as follows:
\begin{equation}
    \mathcal{L}_{CE} = \sum\nolimits_{(q,d)}\sum\nolimits_i^L \log p(id_{d_i} | id_{d_{<i}} , q) , 
\label{eq:retrievalloss}
\end{equation}
where the probability distribution $p(id_{d_i} | id_{d_{<i}} , q)$ is estimated by $\texttt{Decoder}(\texttt{Encoder}(q); id_{d})$. Eventually, the learning of ID assignment and retrieval task is end-to-end trained by
\begin{equation}
    \mathcal{L}_{ASI} = \mathcal{L}_c +  \mathcal{L}_{CE} . 
\end{equation}

Furthermore, we explore multiple implementations of the semantic indexing module. First, following~\citet{yang2023auto}, we implement the indexing module as an MLP-based neural quantization module. Considering that vector quantization has been widely explored to discretize dense vectors in many recent applications, we then implement a differentiable Product Quantization (PQ)-based \cite{zhan2022repconc} indexing module that generates a discrete numeric index of length $ L $ by clustering assignment. We also improve traditional Residual Quantization (RQ) \cite{lee2022autoregressiverq, rajput2024recommenderrq} into the differentiable semantic indexing module. Please refer to Appendix \ref{app:impl} for details of each implementation\footnote{Performance of \ours refer to the neural quantization variant in our experiments, unless otherwise stated.}.

\subsection{Distribution Balancing Criterion: Fostering the Utilization of Indexing Space}
\label{sec:dist_loss}
As aforementioned, one of the key challenges in end-to-end learning of generative retrieval and ID assignment lies in the inefficient and unbalanced utilization of indexing space. Therefore, we propose a novel distributionally balanced criterion for ID assignments to foster the utilization and balance of indexing space during fully end-to-end training via a proposed \textbf{distribution density loss} $\mathcal{L}_{di}$. 

Specifically, $\mathcal{L}_{di}$ focuses on distributional-level properties of the corpus. Formally, we aim to improve the utilization of indexing space without harming retrieval performance as follows: 
\begin{equation}
\begin{aligned}
    \max \mathbb{E}_{d \sim D} \mathbb{E}_{d' \sim D_d' \backslash d} \left[\mathfrak{D}(id_d,id_{d'})\right]  + \mathbb{E}_{q \sim Q} \mathbb{E}_{q' \sim Q_q' \backslash q} & \left[\mathfrak{D}(id_q,id_{q'})\right]  
    \\ 
    & s.t. \min \mathbb{E}_{(q,d^+) \sim C}\left[\mathfrak{D}(id_q,id_{d^+})\right],
\label{eq:max_dist}
\end{aligned}
\end{equation}
where $D$ and $Q$ denote the full observed training corpus for documents or queries, $C$ denotes the corpus of ground-truth positive pairs for training, $D_d'$ and $Q_q'$ denote neighbors of $d$ and $q$ under distance $\mathfrak{D}(\cdot , \cdot)$. Essentially, Eq.\eqref{eq:max_dist} promotes a balanced utilization of indexing space through promoting the exploration of the \textit{entire} space, while achieving an equilibrium to retain the identical matching of relevant query and documents for retrieval performance. 

To efficiently implement Eq.\eqref{eq:max_dist}, we estimate $D_d'$ and $Q_q'$ with a Monte-Carlo batch sample from the full corpus for an arbitrary $d$ and $q$. Since the discrete numeric IDs are equal to a point in the discrete indexing space $\mathbb{N}^{L \times V}$, we could naturally adopt Euclidean distance for $\mathfrak{D}(\cdot , \cdot)$. To overcome the non-differentiable nature of argmax operation when obtaining $id_q$ and $id_d$, we infuse the objective in Eq.\eqref{eq:max_dist} into a differentiable reverse cross-entropy loss \cite{pang2018towards}, which also minimizes the density through maximizing the Kronecker Delta distance of indexes from irrelevant queries or documents. We thereby obtain $\mathcal{L}_{di}$ as follows (take documents as an example): Specifically, $\mathcal{L}_{di}$ focuses on distributional-level properties of the corpus. Formally, we aim to improve the utilization of indexing space without harming retrieval performance as follows: 
\begin{equation}
    \mathcal{L}_{di} = \mathbb{E}_{d' \sim D_d'}\left[\mathfrak{D}(id_d,id_{d'})/\mathfrak{D}_{max}\right]\left[\sum\nolimits_{d' \sim D_d'} CE(id_d,id_{d'})\right],
\label{eq:loss_distance}
\end{equation}
where $CE$ denotes cross-entropy, and we normalize $\mathfrak{D}(d,d')$ to the range of $[0,1]$ by dividing the maximized possible distance within the indexing space $\mathbb{N}^{L \times V}$. Eq.\eqref{eq:loss_distance} jointly captures the balance and density of indexing space through Euclidean and Kronecker Delta distances, fostering the efficient and balanced utilization of indexing space during fully end-to-end training.

\begin{figure}[!t] 
  \centerline{
  \includegraphics[width=\linewidth]{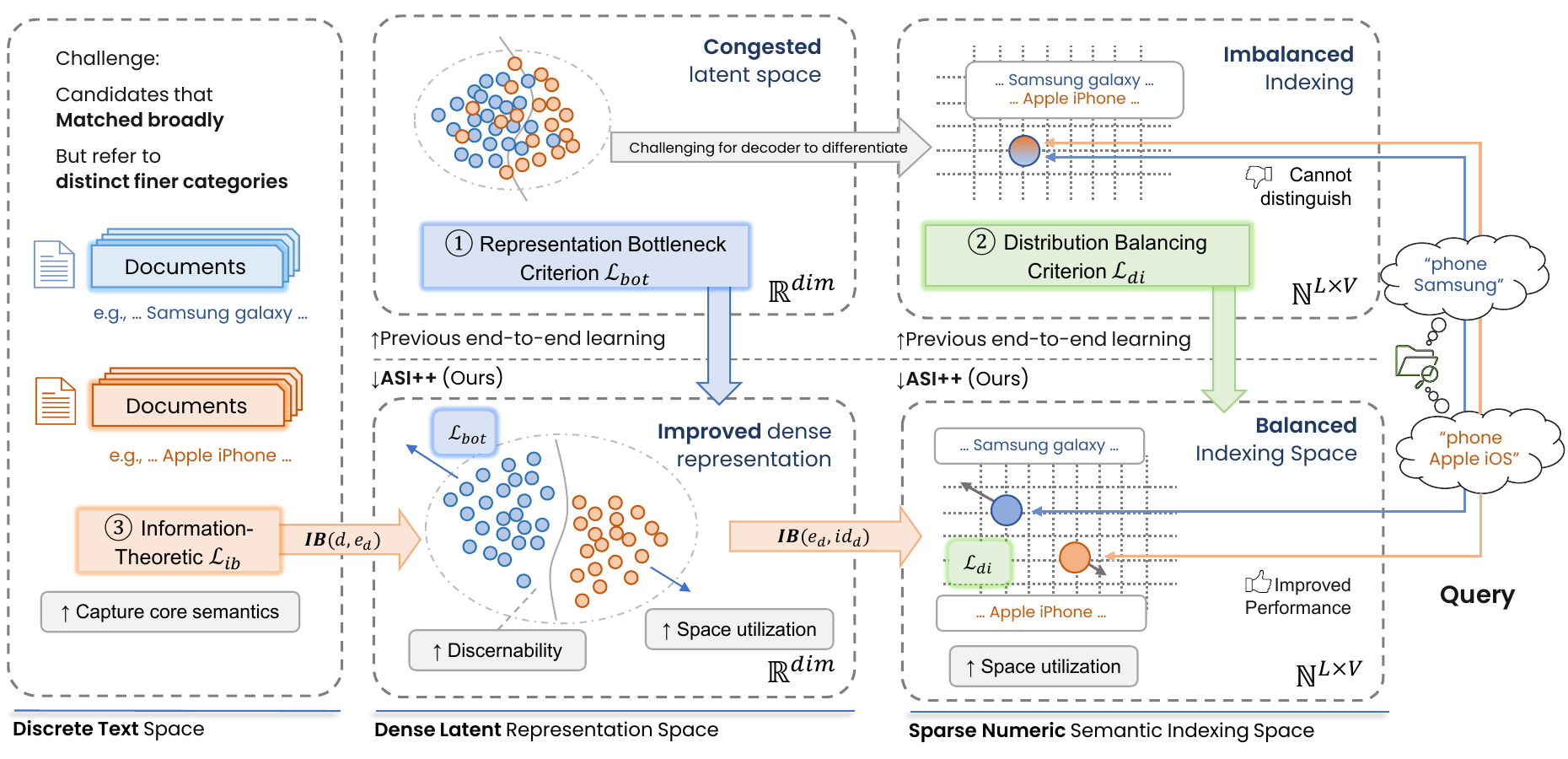}}
  \caption{Summary of main contributions of \ours. (\textit{Upper}) Previous end-to-end learning of GenIR face challenges in efficiently utilizing the dense latent and sparse indexing space. (\textit{Down}) \ours optimizes the distribution of ID distributions, by (\textit{i}) improving the quality and space utilization of dense representations to distinguish similar documents easier with decoder ($\mathcal{L}_{bot}$),  (\textit{ii}) fostering the utilization of indexing space to better match relevant query-documents ($\mathcal{L}_{di}$), and (\textit{iii}) compressing most crucial information into dense bottleneck representations with information-theoretic $\mathcal{L}_{ib}$.}\label{dist-optim}
\end{figure}

\subsection{Representation Bottleneck Criterion: Enhancing the Representation of Dense Bottleneck}
\label{sec:dense_bottleneck}
To improve the utilization and distribution of indexing space, we regularize the training of \ours in Section \ref{sec:dist_loss} through its terminal outputs to improve the learning of the indexing module and decoder in \ours. Revisiting the design of \ours (Figure \ref{fig:model}), it evidently follows a Markov Chain of $<id_q \to q \to e_q>$ and $<id_d \to d \to e_d>$\footnote{Assuming $p(e_x|x) = p(e_x|x, id_x)$, we have $p(x,e_x,id_x)=p(id_x) p(x|id_x) p(e_x|x, id_x)=p(id_x) p(x|id_x) p(e_x|x)$. Here the notation $x$ may refer to query $q$ or document $d$.}, where dense representations from encoder form as a representation bottleneck~\cite{cover1999elements} that is crucial to the IDs learned by \ours. In this subsection, we alternatively focus on enhancing the \textbf{quality of the bottleneck dense representations}.

We consider the quality of dense representation from the following two perspectives: (i) \textbf{Semantic discernability} for relevant and irrelevant ($e_q,e_d$) pairs, which could enforce a good matching of $(id_q, id_{d^+})$ and distinguishing of $(id_q, id_{d^-})$; (ii) \textbf{Good utilization of dense latent space} makes it easier for decoder and indexing module to produce balanced IDs that highly utilize the whole indexing space, following Eq.\eqref{eq:max_dist}. Based on these objectives, we propose a novel loss $\mathcal{L}_{bot}$ as follows\footnote{Here we do not normalize by $\mathfrak{D}_{max}$ in Eq.\eqref{eq:loss_bottleneck} since we have L2-normalized $e_q$ and $e_d$.}:
\begin{equation}
    \mathcal{L}_{bot} \!\!=\!\! - \mathbb{E}_q \!\left[ \log \frac{\mathbf{e}^{(e_q \cdot e_{d^+})}} {\sum\nolimits_{d^-} \mathbf{e}^{(e_q \cdot e_{d^-})}} \right] \!\!+\!\! \gamma \! \left(\mathbb{E}_{d \sim D, d' \sim D_d'}\!\left[\frac{1}{\mathfrak{D}(e_d,e_{d'})}\right] \!\!+\! \mathbb{E}_{q \sim Q, q' \sim Q_q'}\!\left[\frac{1}{\mathfrak{D}(e_q,e_{q'})}\right]\! \right) ,
\label{eq:loss_bottleneck}  
\end{equation}
where $\gamma$ is a hyperparameter for controlling the separation degree of dense latent space, 
and we follow Section \ref{sec:dist_loss} to estimate $D_d'$ and $Q_q'$ with Monte-Carlo samples. The first term of Eq.\eqref{eq:loss_bottleneck} enhances the latent dense representation through contrastive InfoNCE \cite{oord2018representationinfonce} loss, and the second term implements the distribution density objective (Eq.\eqref{eq:max_dist}) on the latent dense representations, enhancing the capability of dense representations.

\subsection{Information Consistency Criterion: Information-Theoretic Semantic Compression}
So far, in Section \ref{sec:dist_loss} and \ref{sec:dense_bottleneck}, we derived objectives for sparse index and dense latent representations on optimizing their space-balanced utilization and quality. 
However, since the neural network is usually non-convex, the optimization directions of the above two objectives may conflict with each other, i.e., each optimizes towards a different (local) optimal solution. Therefore, it is essential to ensure the information consistency of their optimization directions, which we call the information consistency criterion. 
In this section, we still start with Markov Properties of \ours learning: $<id_q \to q \to e_q>$ and $<id_d \to d \to e_d>$ and derive a novel objective $\mathcal{L}_{ib}$ for joint learning of both representations under an \textbf{information-theoretic perspective}, which further gives optimal semantic compression to only retain crucial information for retrieval performance. Specifically, we start deriving $\mathcal{L}_{ib}$ from maximizing the information-bottleneck \cite{tishby2000information} under learning document representations (Here we slightly abuse the notation to refer to the input documents as $x$ for clarity) :
\begin{equation}
    \max I(id_d, e_d) - \beta I(x, e_d).
    \label{eq:general_ib}
\end{equation}

We first derive the variational upper-bond \cite{alemi2016deep} of $I(x, e_d)$ following the non-negativity properties of KL-divergence \cite{cheng2020club}:
\begin{align}
    I(x, e_d) \!&=\! \int p(x, e_d) \log \frac{p(x, e_d)}{p(x)p(e_d)} dx d e_d 
    = \int p(x, e_d) \log p(e_d|x)dxde_d - \int p(e_d) \log p(e_d) de_d  \nonumber \\
    \!&\! \leq \int p(x) p(e_d|x) log \frac{p(e_d|x)}{q(e_d)} dxde_d  = \mathbb{E}_x D_{KL}\left[p(e_d|x)||q(e_d)\right],
    \label{eq:ib_upperbound}
\end{align}

where $q$ is a prior distribution for dense encoding $e_d$. Following existing literature \cite{zhang2022improvinginfobert}, we model $q$ as a Gaussian distribution $\mathcal{N}(\mu, \sigma^2)$, and estimate it with a batch of Monte-Carlo samples \cite{shapiro2003montecarlo}. On maximizing $I(id_d, e_d)$, considering it enforces the bottleneck $e_d$ to accurately capture crucial information for indexing, it has already existed in Eq.\eqref{eq:retrievalloss} as the task-oriented cross-entropy loss, so here we leave this item. 

We thereby obtain the information-theoretic objective $\mathcal{L}_{ib}$ for \ours as follows,  
\begin{equation}
    \mathcal{L}_{ib} = \beta \cdot \mathbb{E}_{d \sim D} D_{KL}\left[p(e_d|d)||q(e_d)\right] .
    \label{eq:ibloss}
\end{equation}

Ultimately, our \ours not only jointly trains both ID assignments and retrieval tasks in an end-to-end manner by $ \mathcal{L}_{ASI} $, but also simultaneously emphasizes efficient and distribution-balanced ID assignments by adding the above objectives. The overall objective for \ours could be thereby expressed as $ \mathcal{L} = \mathcal{L}_{ASI} + \lambda \left( \mathcal{L}_{di} + \mathcal{L}_{bot} + \mathcal{L}_{ib} \right) $, where $\lambda$ is a scaling coefficient. 

\section{Experiments}
\subsection{Experimental Setup}
\paragraph{Datasets and Metrics}
We evaluate the performance of \ours on two document retrieval datasets, including a well-studied public dataset MS MARCO \cite{nguyen2016ms}, and an industrial dataset ADS \cite{yang2023auto}. MS MARCO document ranking is a large-scale collection of queries and web pages for machine reading comprehension. We adopt the official split of the dataset following \cite{yang2023auto}, and apply DocT5Query for query generation \cite{nogueira2019doc2query}. 
\ifpeerreview
\else
ADS \cite{yang2023auto} is a real-world dataset from Bing\footnote{https://www.bing.com} sponsored search consisting of query-ad pairs, where the ads are a concatenation of title and abstract of the target ads to a user's query. 
\fi
We report Recall@K (R@1/5/10) and MRR@10 metrics\footnote{Recall and MRR treat the cases as true positives when the decoder generates the correct ID.}. Since \ours allows a one-to-many mapping during retrieval, we follow \cite{yang2023auto} to report the expectation of random sampling under natural order for a comprehensive evaluation. 
On the ADS dataset, we leverage an offline reward model to measure the macro/micro averaged quality score, i.e., Mi-QS and Ma-QS. Besides, we also report the learned unique ID number, selected high-quality number, and ratio. To prevent retrieving too many documents, we randomly truncate to a maximum of 1000 documents per query. 

\paragraph{Baselines} We compare our proposed \ours with a variety of strong sparse, dense, and generative retrieval baselines. For sparse retrieval, we select the difficult-to-beat BM25 \cite{robertson2009probabilistic}, as well as DocT5Query \cite{nogueira2019doc2query}. For dense retrieval, we select SimCSE \cite{gao2021simcse}, RepBERT \cite{zhan2020repbert}, Sentence-T5 \cite{ni2022sentence}, and GTR \cite{ni2022gtr}. For generative retrieval, we consider DSI \cite{tay2022transformerdsi}, NCI \cite{wang2022neuralnci}, SEAL \cite{bevilacqua2022autoregressiveseal}, DynamicRetriever \cite{zhou2022dynamicretriever}, Ultron \cite{zhou2022ultron}, ASI \cite{yang2023auto}, NOVO \cite{wang2023novo}, and GenRet \cite{sun2024learninggenret}.

\paragraph{Implementation Details}
We adopt Transformers \cite{vaswani2017attention} architecture for the encoder and decoder in \ours, and apply the \texttt{bert-base-uncased} tokenizer for the encoder. For the decoder, we apply a unique set of vocabulary for modeling the indexing space, including 1,024 discrete numeric tokens: $[1,1024]$, and two special tokens for \texttt{bos} and \texttt{eos}. We train the decoder (and its vocab) from scratch\footnote{The decoder in \ours shares a same architecture with the decoder of \texttt{facecook/bart-base}.} and initialize the encoder with a 6-layer pretrained BERT\footnote{Availale at: \texttt{gaunernst/bert-L6-H768-uncased.}}\cite{kenton2019bert}. We train \ours with AMD MI200 series GPUs, with a batch size of 8192, maximum input length of 64, and learning rate of $1e-4$, for a total of 300K steps. We take 10K warmup steps and apply the AdamW \cite{loshchilov2018decoupledadamw} optimizer. For hyperparameters in \ours, we set scaling coefficients $\lambda=0.25$, weight $\gamma=0.05$, IB compression factor $\beta=0.01$ based on preliminary experiments. We follow \cite{yang2023auto} to apply $\alpha=3$ for margin, and $L=4$ for indexing (ID) length. For inference, we apply vanilla beam search (non-constrained), with a beam size of 10. Please refer to Appendix \ref{app:impl} for further details on implementation.

\begin{table}[t]
  \resizebox{\linewidth}{!}{
  \begin{minipage}[b]{0.5\linewidth}
    
\resizebox{\linewidth}{!}{
\begin{tabular}{@{\,\,}l@{\,\,\,}|@{\,\,\,}c@{\,\,\,}|@{\,\,\,}c@{\,\,\,}c@{\,\,\,}c@{\,\,}}
\toprule
Model                                & Params                  & R@1    & R@10  & MRR@10 \\
\midrule
BM25$\dagger$                        & -                       & 39.1 & 69.1 & 48.6 \\
SPLADE$\dagger$                      & -                       & 45.3 & 74.7 & 54.6 \\
\midrule
Sentence-T5$\dagger$                 & 220M                    & 41.8 & 75.4 & 52.8 \\
SimCSE$\ddagger$                     & 110M                    & 28.7	& 73.2 & 43.9 \\
GTR-Base$\dagger$                    & 110M                    & 46.2	& 79.3 & 57.6 \\
\midrule
DSI-Semantic$\dagger$                & 250M                    & 25.7 & 43.6 & 33.9 \\
DSI-Atomic$\dagger$                  & 495M                    & 32.5 & 69.9 & 44.3 \\
NCI$\ddagger$                        & 376M                    & 30.1 & 64.3 & 41.7 \\
SEAL$\ddagger$                       & 139M                    & 28.8 & 68.3 & 40.7 \\
Ultron-URL$\dagger$                  & 248M                    & 29.6 & 67.8 & 40.0 \\
Ultron-PQ$\dagger$                   & 257M                    & 31.6 & 73.1 & 45.4 \\
Ultron-Atomic$\dagger$               & 495M                    & 32.8 & 74.1 & 46.9 \\
GenRet$\dagger$                      & 215M                    & 47.9 & 79.8 & 58.1 \\
NOVO$\dagger$                        & 220M                    & 49.1 & 80.8 & 59.2 \\
ASI                                  & 125M                    & \underline{61.2} & \underline{82.1} & \underline{68.6} \\
ASI (Expectation)                    & 125M                    & 55.0 & 74.1 & 61.8 \\
\midrule
\ours                                & 126M                    & \textbf{61.1} & \textbf{83.4} & \textbf{69.0} \\
\ours (Expectation)                  & 126M                    & 58.2 & 83.3 & 65.8 \\
\bottomrule
\end{tabular}
}

    \caption{Performance on dataset MS MARCO.``$\dagger$'' denotes that the performance is referred from \citet{zhou2022ultron}, \citet{sun2024learninggenret} or \citet{tang2023semanticsedsi} and ``$\ddagger$'' denotes we reproduce by official implementations.}
    \label{tab:marco_perf}
  \end{minipage}
  \qquad
  \begin{minipage}[b]{0.45\linewidth}
    \resizebox*{\linewidth}{!}{
\small
\begin{tabular}{l@{\;\;} | @{\;\;}c@{\;\;}c@{\;\;}c @{\;}|@{\;} c@{\;}c}
\toprule
\multicolumn{6}{l}{\cellcolor[rgb]{ .886,  .937,  .855} Documents in Training \& Validation Set ($\sim$13M Candidates)}  \\
\midrule
Metrics              & R@1      & R@5      & R@10     & \begin{tabular}[c]{@{}c@{}}Mi-QS\\Ma-QS\end{tabular} & D/Q \\
\cmidrule(lr){1-6}
SEAL$^\dagger$                 & 0.0444   & 0.1463   & 0.1998   & 0.5291        & 10   \\ 
\cmidrule(lr){1-6}
SimCSE$^\dagger$               & 0.0934   & 0.2853   & 0.3891   & 0.3122        & 10   \\
\cmidrule(lr){1-6}
Ultron-PQ$^\dagger$          & 0.0281   & 0.0720   & 0.0854   & \begin{tabular}[c]
{@{}c@{}}0.4343\\0.4302\end{tabular} & 10 \\
\cmidrule(lr){1-6}
SimCSE$_\text{ASI}$$^\dagger$ & 0.2768   & 0.4593   & 0.5008   & \begin{tabular}[c]{@{}c@{}}0.5290\\0.5087\end{tabular} & 443  \\
\cmidrule(lr){1-6}
 
SimCSE$_\text{ASI++}$$^\dagger$ & 0.3529   & 0.5032   & 0.5347   & \begin{tabular}[c]{@{}c@{}}0.5181\\0.5066\end{tabular} & 554  \\
\cmidrule(lr){1-6}

ASI          & 0.3952   & 0.6542   & 0.7259   & \begin{tabular}[c]{@{}c@{}}0.4556\\0.4549\end{tabular} & 957 \\
\cmidrule(lr){1-6}
ASI++ (Ours)         & 0.6339   & 0.8385   & 0.8837   &

\begin{tabular}[c]{@{}c@{}}0.4596\\0.4596\end{tabular} & 891 \\
\midrule
\midrule
\multicolumn{6}{l}{\cellcolor[rgb]{ .886,  .937,  .855}Full Documents Collection ($\sim$1B Candidates)}          \\
\midrule

Metrics           & \begin{tabular}[c]{@{}c@{}}ID\\Number\end{tabular}    & Mi-QS    & Ma-QS   & \begin{tabular}[c]{@{}c@{}}Selected\\Number\end{tabular} & \begin{tabular}[c]{@{}c@{}}Selected\\Ratio\end{tabular} \\
\cmidrule(lr){1-6}
ASI           & 3.04 M    & 0.4450        & 0.4450        & 307.19 & 0.3072 \\
\cmidrule(lr){1-6}
ASI++ (Ours)           & 6.89 M      & 0.4471     & 0.4472      & 326.72 & 0.3271 \\

\bottomrule
\end{tabular}
}

    \caption{Performance on dataset ADS. ``$\dagger$'' denotes that the performance is referred from \citet{yang2023auto} and ``$\ddagger$'' denotes our reproduction from own implementation.}
    \label{tab:ads_perf}
  \end{minipage}
  }
\end{table}

\subsection{Results}
\begin{table}[t] 
\centering
\resizebox{0.95\linewidth}{!}{

\tabcolsep0.16 in
\begin{tabular}{l | cc | cc | cc | cc}
\toprule
Metrics    & \multicolumn{2}{c|}{\textbf{Full Valid}} & \multicolumn{2}{c|}{\textbf{Existing}} & \multicolumn{2}{c|}{\textbf{New Content}} & \multicolumn{2}{c}{\textbf{New Semantic}} \\
\midrule
Variants   & ASI & \ours  & ASI & \ours  & ASI & \ours & ASI & \ours       \\
\midrule
\# Queries & 10000  & 10000  & 1661   & 1661   & 8146   & 8194   & 193    & 145         \\
R@1        & 0.3952 & 0.6339 & 0.4218 & 0.6195 & 0.4088 & 0.6451 & 0.0570 & 0.2850      \\
R@5        & 0.6542 & 0.8385 & 0.6865 & 0.8272 & 0.6629 & 0.8470 & 0.2073 & 0.5855      \\
R@10       & 0.7259 & 0.8837 & 0.7583 & 0.8742 & 0.7333 & 0.8905 & 0.2487 & 0.6788      \\
Mi-QS      & 0.4556 & 0.4596 & 0.4675 & 0.4712 & 0.4536 & 0.4577 & 0.4381 & 0.4358      \\
Ma-QS      & 0.4596 & 0.5366 & 0.4666 & 0.4701 & 0.4529 & 0.4580 & 0.4376 & 0.4382      \\
D/Q        & 957    & 891    & 971    & 934    & 955    & 884    & 906    & 806         \\
\bottomrule
\end{tabular}
}
\caption{Performance of \ours under cold start scenario on new documents from ADS.}
\label{tab:new}
\vspace{-10px}
\end{table}

\paragraph{Results on MS MARCO} We first evaluate \ours on a popular public benchmark, MS MARCO document. As illustrated in Table \ref{tab:marco_perf}, \ours significantly outperforms existing sparse, dense, and generative retrieval baselines, including strong competitors like GenRet, NOVO, and ASI. Particularly, \ours gains a significant advantage in R@1 and MRR metrics, demonstrating the superiority of the fully-end-to-end training of \ours. Compared with its predecessor ASI, \ours demonstrates a strong improvement in the expectation of all metrics while also establishing improvements on canonical metrics. These results demonstrate that \ours could still leverage the efficient one-to-many mapping to retrieve high-quality candidates, but significantly remedies the long-tailed and unbalanced distribution of indexing space utilization.

\paragraph{Result on ADS} To test the performance of \ours under real-world large-scaled corpus, we conduct experiments on the ADS dataset collected from a sponsored search engine. We further construct two splits, one is a training set that consists of 13M documents, and another is the full document collection with 1B documents from the sponsored search platform. To mitigate the limitations of qrels\footnote{Since it is challenging to collect all ground-truth documents in a billion-scaled corpus for an arbitrary query.}, we leverage an offline reward model to assess the quality of all retrieved documents. As illustrated in Table \ref{tab:ads_perf}, while SEAL could retrieve high-quality documents, it could only retrieve ten documents under given compute (set topk=10 for beam search in decoder), while pre-processing based Ultron scales poor to large-scaled corpus. In contrast, \ours achieves a high R@K metric, enjoys higher efficiency by returning more high-quality documents, and demonstrates a strong adaptability to real-world, large-scale corpus. On the other hand, compared with ASI, it can be seen that ASI++ learns the document IDs in a more dispersed and balanced manner. It should intuitively have made ID decoding more difficult, but \ours obtains both higher R@K and Quality Scores. These results further illustrate the superiority of \ours's design.

\paragraph{Results on New Documents} We compare \ours against baselines on new documents. Following \cite{yang2023auto}, we split queries in ADS dataset into three new categories, including `Existing' (documents are seen during training), `New Content' (documents are unseen but their IDs are seen), and `New Semantic' (New documents with OOD IDs). As shown in Table \ref{tab:new}, \ours largely outperforms the baseline, especially under the most challenging `New Semantic' scheme. We conjecture the novel criteria proposed in \ours (dense representation criterion, information-theoretic optimization) play a vital role in capturing the core semantics of documents, and thus improve the capability and generalizability of the semantic indexing process. \label{sec:new_doc}

\begin{figure}[t]
  \begin{minipage}[b]{0.45\linewidth}
  \centering
  \resizebox{\linewidth}{!}{\includegraphics[]{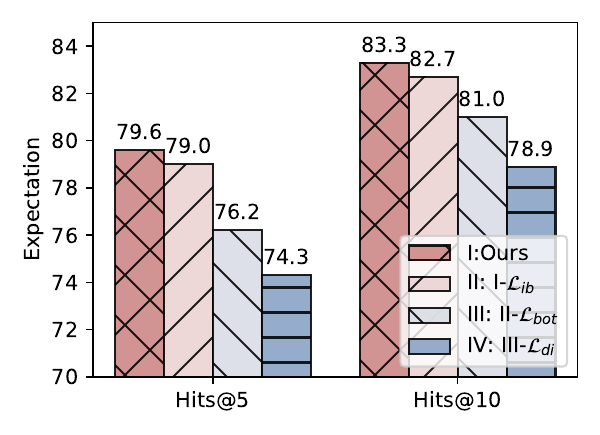}}
    \vspace{-15px}
    \captionof{figure}{Ablation of \ours on MS MARCO. We report the expectation of R@5/10.}
    \label{fig:aba}
  \end{minipage}
  \qquad
  \begin{minipage}[b]{0.48\linewidth}
  \resizebox{\linewidth}{!}{
\centering
\begin{tabular}{@{\,\,}l@{\,\,\,}|@{\,\,\,}c@{\,\,\,}|@{\,\,\,}c@{\,\,\,}c@{\,\,\,}c@{\,\,\,}c@{\,\,}}
\toprule
Model          & Impl.   & R@1    & R@5    & R@10  & MRR@10 \\
\midrule
\ours         & PQ    & 57.2 & 76.1 & 80.6 & 65.4 \\
              & RQ    & 55.1 & 75.9 & 80.3 & 59.7 \\
              & MLP   & 61.1 & 79.6 & 83.4 & 69.0 \\
\midrule
\ours          & PQ   & 54.1 & 76.1 & 80.6 & 62.0 \\
(Expectation)  & RQ   & 51.1 & 75.8 & 80.3 & 59.7 \\
               & MLP  & 58.2 & 79.4 & 83.3 & 65.8 \\
\bottomrule
\end{tabular}

}
    \caption{Performance of different variants of \ours on MS MARCO. We explore multiple implementation of the semantic indexing module: product quantization (PQ), residual quantization (RQ), and MLP-based neural quantization.}
    \label{tab:indexing}
  \end{minipage}
\end{figure}

\paragraph{Variants} In this section, we explore the performance of different implementations of the semantic indexing module proposed in \ours, including product quantization (PQ), residual quantization (RQ), and MLP-based neural quantization (MLP). Results are presented in Table \ref{tab:indexing}. From the results, we observe that the MLP-based neural quantization module generally outperforms product quantization, then residual quantization. We conjecture that the MLP module is more expressive and fits our end-to-end learning setting, whose training dynamics might potentially require major adjustments of indexes, whereas quantization-based methods may better fit with well-established dense representations \cite{zhan2022repconc}. However, despite these differences, the general performance of \ours varying implementations is considerably competent, indicating the effectiveness of \ours's design.

\paragraph{Ablation Study}
To study the effect of each proposed module in \ours, we conduct an ablation study on the MS MARCO dataset. Results are illustrated in Figure \ref{fig:aba}. We first remove the information consistency criterion, i.e., the objective $\mathcal{L}_{ib}$, and the performance drops across both metrics. We then remove the representation bottleneck criterion $\mathcal{L}_{bot}$ and next the distribution balancing criterion $\mathcal{L}_{di}$, and the performance drops progressively. These results indicate that the proposed components of \ours bring complementary effects to improving the quality of document indexing and retrieval.

\section{Analysis}
\vspace{-5pt}
\subsection{Case Study}
\label{sec:case}
\begin{figure}[!t]
  \centerline{
  \includegraphics[width=\linewidth]{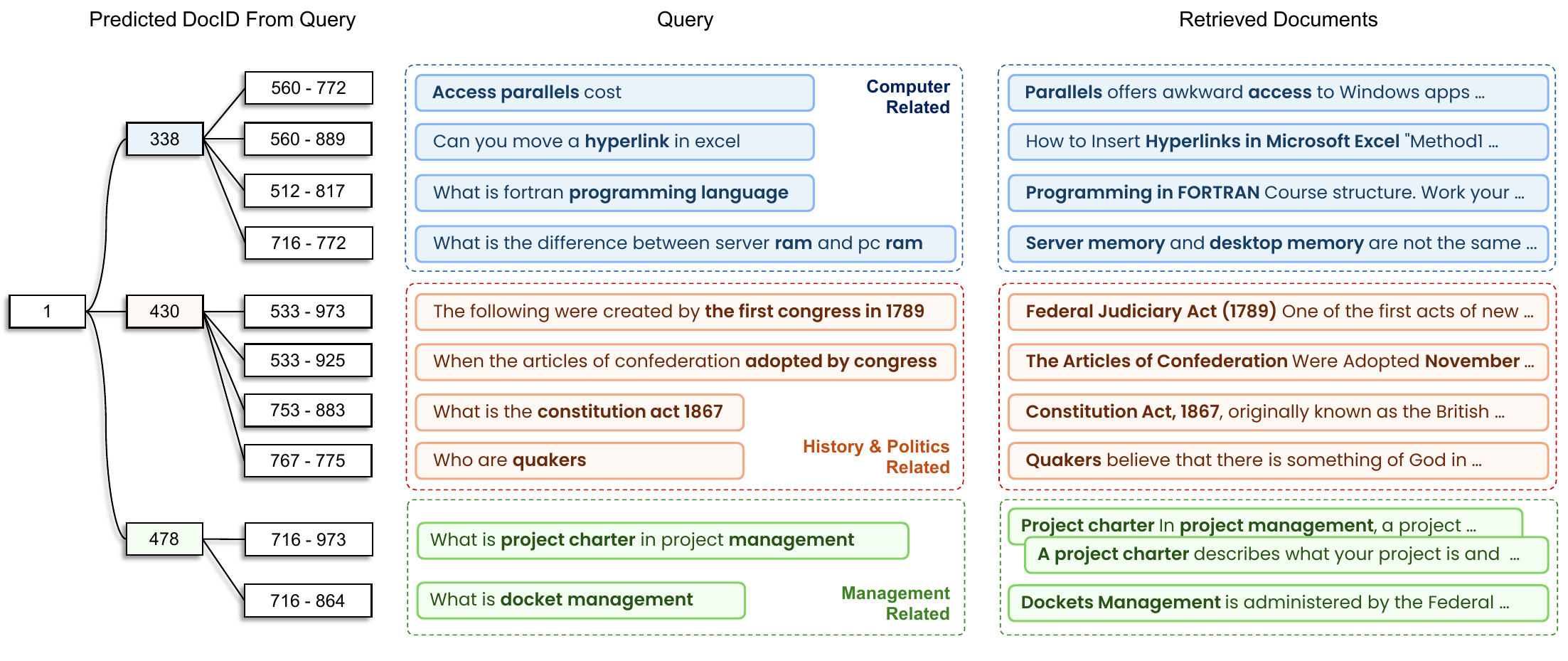}}
  \caption{Qualitative case study on the interpretability of IDs generated from \ours on MS MARCO.}
  \label{fig:case}
  \vspace{-5px}
\end{figure}

In this section, we present a case study on the interpretability of document indexes generated through our proposed \ours. We first illustrate the indexing space via a prefix tree in Figure \ref{fig:case}. We observe that even unlike previous works (e.g. pre-processed IDs from hierarchical clustering \cite{tay2022transformerdsi, wang2022neuralnci}) that \textit{no heuristic priors are given}, the indexing space learned fully end-to-end by \ours \textit{naturally} forms a hierarchical semantic structure. For example, take \texttt{1} as the first digit, for the digit at 2$^\text{nd}$ position, queries or documents with index \texttt{336} correlates with Computer-related topics, \texttt{430} for History and Politics, \texttt{478} for management areas, etc. We also observe that the 3$^\text{rd}$ and 4$^\text{th}$ position capturing finer-grained semantics, e.g., prefix \texttt{1-430-533} specifically correlates to topics that related to the first congress of the US, prefix \texttt{1-338-560} for applications on the Windows OS. Case \texttt{1-478-716-893} demonstrates the one-to-many mapping capability of \ours to retrieve multiple semantically close documents for a given query. This hierarchical property on semantics of the indexing space to \ours shed light on its interpretation capability.

\subsection{Qualitative Study of Indexing Space Utilization}
\begin{wrapfigure}{r}{6.5cm}
\vspace{-30px}
\resizebox{6cm}{!}{
  \includegraphics[width=6cm]{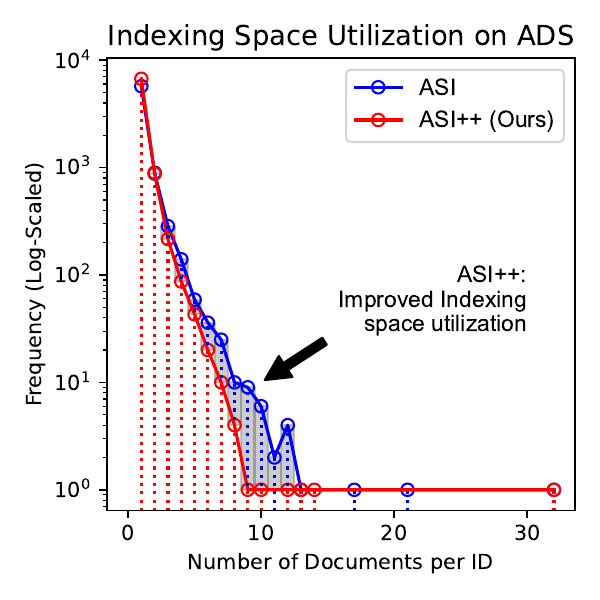}
}
\vspace{-10px}
\caption{Comparison of the indexing space utilization of \ours and ASI on ADS dataset.}
\label{fig:density}
\vspace{-20px}
\end{wrapfigure}

To further study how \ours improves the utilization of indexing space, we compare \ours against ASI on their ID assignments, from the ADS dataset. Specifically, we sample 10,000 unique documents in the ADS dataset, and capture the statistics on how frequently a same DocID is occupied by different documents. As illustrated in Figure \ref{fig:density}, \ours improves the utilization of indexing space by assigning more unique identifiers to documents (the shadowed region). Intuitively, this reflects a more thorough exploration of indexing space, and may potentially generalize better to new documents (since they could thereby occupy novel IDs), which corroborates with our findings in Section \ref{sec:new_doc}.

\section{Related Work}
\paragraph{Dense Retrieval}

Dense retrieval leverages neural models to learn document and query representations as embeddings, then achieve retrieval through matching query and candidate embeddings with dot product or cosine similarity \cite{karpukhin2020dense}. Numerous methods have been introduced to improve dense retrieval, including hard negative mining \cite{xiong2020approximate, qu2021rocketqa}, pre-training \cite{ni2022sentence, ni2021largedual}, and knowledge distillation \cite{hofstatter2021efficientlykd}. However, the high dimensionality and demand on vector searching bring significant memory and compute constraints to dense retrieval retrievals. To overcome this limitation, a series of works explored quantization, including product quantization \cite{chen2020differentiablepq, zhan2022repconc}, residual quantization \cite{niu2023residual}, but their performance is generally inferior to original dense models.

\paragraph{Generative Retrieval}

As a new paradigm of retrieval, generate retrieval leverage Transformer's memory \cite{tang2023semanticsedsi} and generate an index auto-regressively, given a query of the document. It is believed that compared to dense retrieval, the generative retrieval world better unleashes the power of base LMs due to its closer nature to their pre-trained tasks \cite{bevilacqua2022autoregressiveseal}. A large strand of works explores the indexing space in generative retrieval, including leveraging unique atomic integer~\cite{zhou2022ultron, tay2022transformerdsi}, n-grams~\cite{bevilacqua2022autoregressiveseal, tang2023semanticsedsi, li2023multiviewminder, chen2022corpusbrain, chen2023unified}, natural language sentences~\cite{de2020autoregressiveentity, chen2022corpusbrain, chen2023unified}, hierarchical clustering~\cite{tay2022transformerdsi,wang2022neuralnci, zhuang2022bridgingdsiqg}, quantization \cite{rajput2024recommenderrq}, and looking up codebooks \cite{sun2024learninggenret, yang2023auto}. 
One significant constraint of generative retrieval lay in the demand on pre-processing of document IDs for training \cite{tang2023semanticsedsi, wang2022neuralnci, zhou2022ultron, bevilacqua2022autoregressiveseal}. To mitigate this limitation, \citet{sun2024learninggenret} proposes a progressive end-to-end training, while \citet{yang2023auto} achieves a fully end-to-end training. Compared to prior approaches, \ours achieves fully end-to-end training and efficient retrieval, and introduces multiple novel criterions to foster better space utilization during training.

\section{Conclusion}
We propose \ours, a novel fully end-to-end generative retriever. \ours inherits the essence of ASI's design: fully end-to-end training, efficient one-to-many mapping, and substantial improvements through optimizing the distributional balance of indexing space utilization. Specifically, in \ours, we derive multiple novel criterions, including distributionally balanced criterion, and representation bottleneck criterion, and connect them cohesively through an informational-theoretic perspective. Through jointly optimizing these criterions, \ours achieves a significant improvement in utilizing the indexing space, demonstrating competitive performance against strong baselines on both public and large-scaled industrial retrieval tasks. We further conduct a comprehensive analysis to demonstrate the interpretability of indexes produced by \ours, and \ours's advantage under cold-start scenarios.

\bibliographystyle{unsrtnat}
\bibliography{reference}

\newpage
{\LARGE \textbf{Appendix}}
\appendix
\section{Limitations and Broader Impacts}
\paragraph{Limitations} \ours leverages a fully end-to-end training scheme and utilizes a sparse numeric indexing space for generating IDs for queries and documents. This design brings multiple benefits, such as efficiency in decoding, better representation of semantic relevance, and generalization capability to multilingual scenarios. While we do observe hierarchical semantic information between different indexes in Section \ref{sec:case}, an ID could be hard to comprehend standalone by humans, compared with n-grams of natural language sequence-based identifiers, which might still pose some challenges in understanding specific cases. The second potential limitation of \ours lies in the ranking capability of \ours. Currently, while our one-to-many mapping proves to be efficient computationally when retrieving documents, we neglect the difference within documents occupying the same ID during retrieval. The combination of \ours with ranking would be a promising direction. Finally, for a fair comparison with prior works, the training of \ours only leverages in-batch negatives. It would be indeed promising to explore negative mining strategies to improve \ours in the future.

\paragraph{Broader Impacts} 
Designed as a fully end-to-end generative retriever, our \ours is efficient to train and performs promisingly against strong dense, and generative retrieval baselines under real-world, large-scale data. We envision \ours being applied to real-world applications, e.g., information retrieval, recommendation, and web search systems, as a promising new paradigm of search. It may also potentially bring complementary effects by integrating with Large language models, under popular use cases such as retrieval-augmented generation, to reduce hallucinations and provide newer knowledge by retrieving higher-quality corresponding documents. Technically, we propose three novel criterion for training fully end-to-end generative retrieval models, and explore various implementations of semantic indexing modules, aiming at a new focus towards distribution-level optimization on document indexing. These novel findings may bring fresh insights into understanding and further improving generative retrieval systems.

\section{Implementation Details}
\label{app:impl}
\subsection{Architecture of Semantic Indexing Module}
In this work, we explore multiple implementations of the semantic indexing module in \ours, including product quantization, residual quantization, and an MLP-based neural quantization. In this section, we introduce each of their designs and implementations in detail.

\paragraph{Product Quantization} As an effective vector quantization approach to obtain sparse indexes from high-dimensional dense vectors, product quantization has been widely applied to dense retrieval methods. In \ours, we adapt a strong PQ implementation from RepCONC \cite{zhan2022repconc}. RepCONC\footnote{RepCONC: \url{https://github.com/jingtaozhan/RepCONC}} is a PQ method that is trained jointly with the retrieval task. It contains a ranking-oriented loss and a PQ loss (MSE between quantized vectors and raw dense vectors). We now briefly introduce the integration of RepCONC as the semantic indexing module of \ours.

Denote $e_x$ and $e^q_x$ the raw and quantized embedding of input $x$ (query or documents). To learn optimal quantization, we first add the quantization MSE loss $||e_x-e^q_x||_2$ into our total objective. We replace the vanilla ranking-oriented loss of RepCONC with $\mathcal{L}_c$ in \ours, and the probability $p(id_x|e_x)$ is defined as Sinkhorn distance \cite{cuturi2013sinkhorn} to each clustering centroid\footnote{The distance is directly derived through Sinkhorn \cite{knight2008sinkhorn} algorithm, please refer to \cite{zhan2022repconc} for more details.}. To model the indexing space of $\mathbb{N}^{L \times V}$, we set the number of clustering groups and centroids in PQ to 4 and 256, respectively. We set $\epsilon=0.003$ and  $iters=100$ for the Sinkhorn algorithm in RepCONC, following its suggested implementation \cite{zhan2022repconc}.

\paragraph{Residual Quantization} We explore another quantization approach - residual quantization (RQ) \cite{rajput2024recommenderrq, zeghidour2021soundstream, vasuki2006reviewquant} in \ours. We maintain 4 codebooks in \ours to generate a 4-lengthed sparse index. Starting with the dense representation $e_x$, we lookup the first codebook and obtain the closest vector $e^1_x$. We take the index of the closest vector as the first position of the sparse index and obtain the residual error $r^1_x = e^1_x - e_x$. We continue to lookup the next index with $r^1_x$, obtain the closest vector $e^2_x$ and residual error $r^2_x = e^2_x - r^1_x$. The process is repeated two more times, after which we obtain 4 sparse indexes for input $x$, and the quantized representation $e^q_x := e^1_x + e^2_x + e^3_x + e^4_x$. 
To jointly train RQ with \ours, we add the quantization MSE loss $||e_x-e^q_x||_2$ into our total objective, and the probability $p(id_x|e_x)$ is similarly defined as Sinkhorn distance to each vector in the codebook (for each position)\footnote{We similarly leverage the implementation of Sinkhorn algorithm in the implementation of \cite{zhan2022repconc}, except that we replace the original PQ vectors and centroids to RQ vectors and centroids.}.

\paragraph{MLP-based Neural Quantization} We finally consider a lightweight and efficient implementation of the semantic indexing module. Specifically, we employ a dense neural network $f$ (MLP with Softmax activations): $\mathbb{R}^d \to \mathbb{R}^{|V|}$, to obtain $p(id_{x_i}|e_x)$. We employ a separate network $f$ for each of 4 positions and adopt a dropout of 0.2 during training. Compared to autoregressive modules (e.g. Transformer), this lightweight design achieves $\mathcal{O}(1)$ inference speed, which is desirable for processing large-scale document corpora.

\subsection{Detailed Configuration}
In this section, we elaborate on further details regarding the implementation of \ours. We list all hyperparameters of \ours in Table \ref{tab:config}, and statistics of datasets\footnote{Link to MS MARCO: \url{https://microsoft.github.io/msmarco/}} in Table \ref{tab:statistics}. Specifically, during the first training epoch, we linearly increase the scaling coefficient $\lambda$ from a small initial value $0.01$ to its defined value. This strategy empirically fosters a desirable initialization of indexing space for further end-to-end training, preventing it from trapping into local optima. 

\begin{table}[h]
\renewcommand{\arraystretch}{1.1}
\centering
\resizebox{0.95\linewidth}{!}{
\begin{tabular}{ll|ll}
\toprule
\textbf{Param}  & \textbf{Value}  & \textbf{Param} & \textbf{Value}  \\
\midrule
Encoder vocab       & \texttt{bert-base-uncased}    & Index length & 4 \\

Batch size & 8192 & Steps & 300,000 \\

Learning rate & 1$e$-4 & Warmup steps & 10,000 \\

Optimizer & AdamW & Adam betas & (0.9, 0.999) \\ 

Scaling coefficient $\lambda$ & 0.25 & Weight $\gamma$ & 0.05 \\

IB factor $\beta$ & 0.01 & Margin $\alpha$ & 3 \\

Decoding beam size & 10 & Compute & MI200 / V100\\

Max length & 64 & Model dim & 768 \\

Indexing range & \begin{tabular}[c]{@{}l@{}}$[1,256], [257,512],$\\$ [513,768], [769,1024]$\end{tabular} & Decoder vocab size & \begin{tabular}[c]{@{}l@{}}1026 \\($[1,1024]$+\texttt{bos}+\texttt{eos})\end{tabular}  \\

\midrule

(PQ) Epsilon & 0.003 & (PQ) Iteration & 100 \\

\bottomrule
\end{tabular}
}
\caption{List of detailed configurations and hyperparameters of \ours}
\label{tab:config}
\end{table}

\begin{table}[h]
\renewcommand{\arraystretch}{1.1}
\centering
\begin{tabular}{lrrr}
\toprule
\textbf{Datasets}  & ADS  & MSMARCO & ADS-Full  \\
\midrule
Train     & 50M   & 367K   & -  \\
Expansion & -     & 32M    & -    \\
Valid     & 10K   & 5.2K   & -    \\
\# Docs   & 13.1M & 3.2M   & $\sim$1B \\
\bottomrule
\end{tabular}
\caption{Statistics of Datasets}
\label{tab:statistics}
\end{table}

\section{Extended Case Study}
\begin{table*}[ht]
\renewcommand{\arraystretch}{1.1}
\centering
%
\small
\begin{tabular}{p{\linewidth}}
\toprule
Docid 1: 24,426,703,876 (\textbf{graduate certificate})                                                                       \\ 
\midrule
\textbf{graduate certificate} in sustainable tourism asu online tourism is an enduringly popular industry however its ...                              \\
\textbf{graduate certificate} in global development asu online arizona state university's graduate certificate  ...                          \\
social entrepreneurship \textbf{graduate certificate} asu online earning a social entrepreneurship and community...                  \\
scu \textbf{graduate certificate} in business 2022 n the \textbf{graduate certificate} in business provides ...                                 \\

\midrule
Docid 2: 227,335,690,792 (\textbf{under armour football})                                                                     \\
\midrule
\textbf{under armour football} apparel for sale new and used on sidelineswap ...             \\
\textbf{under armour football} protective gear curbside pickup available at dick's find what you are looking for n tell ...                                  \\
\textbf{under armour football} gear curbside pickup available at dick's ...                       \\
\textbf{under armour football} game pants for sale new and used on sidelineswap ...                            \\

\midrule
Docid 3: 47,276,709,921 (\textbf{volkswagen eos air filters})                                                                           \\
\midrule
\textbf{volkswagen eos air filters} from 12 carparts com k n engine air filter high performance premium n k n engine ...                                                      \\
2015 \textbf{volkswagen eos air filters} from 11 carparts com n get the best deals on an aftermarket  ...                                                                                  \\
\textbf{volkswagen eos cabin air filters} from 7 carparts com k n premium cabin air filter high ...                                                      \\
2008 \textbf{volkswagen eos cabin air filters} from 7 carparts com k n premium cabin air filter high performance ...           \\

\midrule
Docid 4: 47,367,564,984 (\textbf{volkswagen golf})                                                                          \\
\midrule
2016 \textbf{volkswagen golf} sportwagen abs speed sensors from 21 carparts com warning this product can ...                       \\
buy a 2002 \textbf{volkswagen golf} air temperature sensor at discount prices choose top quality brands beck arnley ...                                 \\
buy a 1998 \textbf{volkswagen golf} camshaft position sensor at discount prices choose top quality brands bosch ...                                                     \\

\midrule
Docid 5: 143,335,589,939 (\textbf{walking shoes \& hush puppies})                                                                          \\
\midrule
comfortable \textbf{walking shoes} for men women \textbf{hush puppies} ...                       \\
shop comfortable \textbf{walking shoes hush puppies} free shipping on orders over 49 n and ...                                 \\
men's \textbf{walking shoes hush puppies} your first order you ll be n arrivals and promotions ...    \\
shop best casual \textbf{walking shoes hush puppies} ... \\

\bottomrule
\end{tabular}%
\vspace{5px}
\caption{Five docid cases of \ours on ADS dataset.}
\label{apdx:tab:case_ads}
\end{table*}


\begin{table*}[t]
\renewcommand{\arraystretch}{1.1}
\centering
%
\small
\begin{tabular}{p{\linewidth}}
\toprule
Docid 1: 69,392,525,814 (\textbf{eating disorder})                                                                       \\ 
\midrule
\textbf{Eating Disorders Eating Disorders} Kathleen N. Franco, MDErin H. Sieke Leah Dickstein, MDTatiana  ... \\
\textbf{Eating disorder} From Wikipedia, the free encyclopedia navigation search Eating disorders Specialty ...\\

\midrule
Docid 2: 157,317,562,798 (\textbf{anxiety})                                                                     \\
\midrule
How \textbf{Anxiety }Influences Your Health (INFOGRAPHIC) No one likes to experience bouts of stress or anxiety ...                            \\
What is\textbf{ Anxiety}? \"What is Anxiety? By: Tanja Jovanovic, Ph. D. Consulting Editor, ... \\
\textbf{Anxiety} Symptoms (including Anxiety Attacks, Disorder, and Panic Signs and Symptoms) ... \\

\midrule
Docid 3: 26,465,671,819 (\textbf{ideal retirement age})                                                                           \\
\midrule
The New \textbf{Ideal Retirement Age}: 67 The age the typical worker expects to retire is no longer ...                                                      \\
The \textbf{Ideal Retirement Age} (i Stock Photo)Retirement at age 65 is no longer ...           \\

\midrule
Docid 4: 150,417,648,798 (\textbf{the demand and supply curve})                                                                          \\
\midrule
What Happens to the Equilibrium Price When Quantity of \textbf{Supply \& Demand Shifts} Upward? Related ...                       \\
What does it mean when the \textbf{supply curve shifts} to the right? Answers.com ...                                 \\
\textbf{Demand curve} From Wikipedia, the free encyclopedia An example of \textbf{a demand curve shifting} ...                    \\

\midrule
Docid 5: 64,499,687,1021 (\textbf{stock market opening time})                                                                          \\
\midrule
What \textbf{time} does the \textbf{Stock Market open}? The stock market opens at 9:00 am and close ...                       \\
What \textbf{Time} Does The \textbf{Stock Market Open And Close}? May 11, 2016, 01:54:48 PM EDT By NASDAQ.com  ...                                 \\

\bottomrule
\end{tabular}%
%
\vspace{5px}
\caption{Five docid cases of \ours on MS MARCO dataset.}
\label{apdx:tab:case_marco}
\end{table*}

We provide additional cases of DocIDs generated by \ours in Table \ref{apdx:tab:case_ads} and \ref{apdx:tab:case_marco}.

\end{document}